\documentclass[11pt]{article}

\pdfoutput=1
\usepackage[utf8]{inputenc}
\usepackage{amsmath,amsfonts,amssymb}
\usepackage{amsthm}
\usepackage{latexsym}
\usepackage[noadjust]{cite}
\usepackage{graphicx}
\usepackage{xcolor}
\usepackage{dirtytalk}
\usepackage{mathtools}
\usepackage[margin=1in]{geometry}
\usepackage{thm-restate}
\usepackage{enumitem}
\usepackage[linesnumbered,ruled]{algorithm2e}
\usepackage[colorlinks=true, allcolors=blue]{hyperref}
\usepackage[nameinlink,capitalise]{cleveref}
\hypersetup{
    citecolor={violet}
}

\newtheorem{theorem}{Theorem}[section]
\newtheorem{corollary}[theorem]{Corollary}

\newtheorem{claim}[theorem]{Claim}
\theoremstyle{definition}
\newtheorem{definition}[theorem]{Definition}
\newtheorem{remark}[theorem]{Remark}
\newtheorem{fact}[theorem]{Fact}

\newcommand{\Z}{\mathbb{Z}}

\newcommand{\zo}{\{0, 1\}}

\newcommand{\eps}{\epsilon}

\newcommand{\cut}{\text{cut}}

\newcommand{\minkcut}{\mathbf{\Phi}}

\newcommand{\polylog}{\mathrm{polylog}}

\newif\ifdraft

\drafttrue

\title{Near-optimal Hypergraph Sparsification in Insertion-only and Bounded-deletion Streams}
\date{\today}
\author{Sanjeev Khanna\thanks{School of Engineering and Applied Sciences, University of Pennsylvania, Philadelphia, PA. Email: {\tt sanjeev@cis.upenn.edu}. Supported in part by NSF award CCF-2402284 and AFOSR award FA9550-25-1-0107.} \and Aaron (Louie) Putterman\thanks{School of Engineering and Applied Sciences, Harvard University, Cambridge, Massachusetts, USA. Supported in part by the Simons Investigator Awards of Madhu Sudan and Salil Vadhan, NSF Award CCF 2152413 and AFOSR award FA9550-25-1-0112. Email: \texttt{aputterman@g.harvard.edu}.} \and Madhu Sudan\thanks{School of Engineering and Applied Sciences, Harvard University, Cambridge, Massachusetts, USA. Supported in part by a Simons Investigator Award, NSF Award CCF 2152413 and AFOSR award FA9550-25-1-0112. Email: \texttt{madhu@cs.harvard.edu}.}}

\begin{document}

\maketitle

\begin{abstract}
We study the problem of constructing hypergraph cut sparsifiers in the streaming model where a hypergraph on $n$ vertices is revealed either via an arbitrary sequence of hyperedge insertions alone ({\em insertion-only} streaming model) or via an arbitrary sequence of hyperedge insertions and deletions ({\em dynamic} streaming model). For any $\epsilon \in (0,1)$, a $(1 \pm \epsilon)$ hypergraph cut-sparsifier of a hypergraph $H$ is a reweighted subgraph $H'$ whose cut values approximate those of $H$ to within a $(1 \pm \epsilon)$ factor. Prior work shows that in the static setting, one can construct a $(1 \pm \epsilon)$ hypergraph cut-sparsifier using $\tilde{O}(nr/\epsilon^2)$ bits of space [Chen-Khanna-Nagda FOCS 2020], and in the setting of dynamic streams using $\tilde{O}(nr\log m/\epsilon^2)$ bits of space [Khanna-Putterman-Sudan FOCS 2024]; here the $\tilde{O}$ notation hides terms that are polylogarithmic in $n$, and we use $m$ to denote the total number of hyperedges in the hypergraph. Up until now, the best known space complexity for insertion-only streams has been the same as that for the dynamic streams. This naturally poses the question of understanding the complexity of hypergraph sparsification in insertion-only streams.

Perhaps surprisingly, in this work we show that in \emph{insertion-only} streams, a $(1 \pm \epsilon)$ cut-sparsifier can be computed in $\tilde{O}(nr/\epsilon^2)$ bits of space, \emph{matching the complexity} of the static setting.
As a consequence, this also establishes an $\Omega(\log m)$ factor separation between the space complexity of hypergraph cut sparsification in insertion-only streams and dynamic streams, as the latter is provably known to require $\Omega(nr \log m)$ bits of space. To better explain this gap, we then show a more general result: namely, if the stream has at most $k$ hyperedge deletions then $\tilde{O}(n r \log k/\epsilon^2)$ bits of space suffice for hypergraph cut sparsification.
Thus the space complexity 
smoothly interpolates between the insertion-only regime ($k=0$) and the fully dynamic regime ($k=m$). 
Our algorithmic results are driven by a key technical insight: once sufficiently many hyperedges have been inserted into the stream (relative to the number of allowed deletions), we can significantly reduce the underlying hypergraph by size by {\em irrevocably} contracting large subsets of vertices.

Finally, we complement this result with an essentially matching lower bound of $\Omega(n r \log(k/n))$ bits, thus providing essentially a tight characterization of the space complexity for hypergraph cut-sparsification across a spectrum of streaming models.

\end{abstract}
\pagenumbering{gobble}

\pagebreak

\tableofcontents

\pagebreak

\pagenumbering{arabic}

\section{Introduction}

In this work, we continue the line of study on designing \emph{hypergraph cut sparsifiers} in the streaming model. 

Recall that a hypergraph, denoted by $H = (V, E)$, is given by a set of hyperedges $E$ where each hyperedge $e \in E$ is an arbitrary subset of the vertices $e \subseteq V$. In the \emph{streaming model}, the hyperedges are revealed in a sequence of steps, where each step consists of either an insertion of a new hyperedge, or the deletion of an existing hyperedge from the hypergraph. At the end of the stream, our goal is to return a \emph{sparsifier} of the resulting hypergraph. Recall that for a hypergraph $H$, and $\eps \in (0,1 )$, a $(1 \pm \eps)$ hypergraph cut-sparsifier of $H$ is a \emph{re-weighted sub-hypergraph} $\hat{H}$ such that \emph{for every} cut $S \subseteq V$, 
\[
\cut_{\hat{H}}(S) \in (1 \pm \eps) \cut_H(S).
\]
Here, $\cut_H(S)$ denotes the total weight of hyperedges that cross from $S$ to $\bar{S}$ (i.e., the set of hyperedges $\{e \in E: e \cap S \neq \emptyset \wedge e \cap \bar{S} \neq \emptyset\}$). 
Cuts in hypergraphs capture a variety of applications, ranging from scientific computing on sparse matrices \cite{BDKS16}, to clustering and machine learning \cite{YNYNLT19, ZHS06}, and to modeling transistors and other circuitry elements in circuit design \cite{AK95, DLaw73}. In each of these applications, the ability to sparsify the hypergraph while still preserving cut-sizes is a key building block, as these dramatically decrease the memory footprint (and complexity) of the networks being optimized. 
In fact, while an arbitrary hypergraph may have as many as $2^n$ distinct hyperedges, the work of Chen, Khanna, and Nagda \cite{CKN20} showed that $(1 \pm \eps)$ hypergraph cut-sparsifiers can be constructed using only $\widetilde{O}(n / \eps^2)$ re-weighted hyperedges, a potentially exponential size saving in the bit complexity of the object. 

This dramatic reduction in the complexity of the hypergraph while still preserving key properties has also prompted a line of research into the feasibility of sparsifying hypergraphs in the \emph{streaming} model of computation. Here, the hyperedges are presented one at a time, with each hyperedge either being inserted into or deleted from the hypergraph constructed so far. In the streaming model, the goal is to compute some function of the hypergraph (in our case, to construct a \emph{sparsifier} of the hypergraph) after the final update in the stream has arrived, while using as few bits of memory as possible in each intermediate step of processing the stream. As mentioned above, computing sparsifiers of a hypergraph in a stream is valuable as it immediately yields itself to streaming algorithms for any problem that relies \emph{only} on a sparsifier (for instance, computing cut sizes, flows, many clustering objectives, and more), and as such has seen study in several papers \cite{GMT15, CKN20, KPS24d}. 

Most recently, the work of Khanna, Putterman, and Sudan \cite{KPS24d} studied the space complexity of of hypergraph cut-sparsification in the setting of {\em dynamic streams} where a hypergraph is revealed via an arbitrary sequence of hyperedge insertions and deletions. They showed an upper bound of $\widetilde{O}(nr \log(m) / \eps^2)$\footnote{Here $\widetilde{O}(\cdot)$ hides $\mathrm{polylog}(\cdot)$ factors. Importantly, in hypergraphs $m$ may be as large as $2^n$, and thus $\log(m)$ can potentially be as large $n$. Hence $\widetilde{O}(\cdot)$ does not hide factors of $\log(m)$.} bits for computing $(1 \pm \eps)$ sparsifiers in dynamic streams, and also established a nearly-matching lower bound of $\Omega(nr \log(m))$ bits, where $n$ is the number of vertices, $r$ is the maximum size of any hyperedge, and $m$ is the number of hyperedges. In particular, because dynamic streaming is a strictly harder setting than insertion-only streams, this also implies an $\widetilde{O}(nr \log(m) / \eps^2)$ bit upper bound for the complexity of constructing hypergraph cut-sparsifiers in the insertion-only model. 

For comparison, in the static setting (i.e., when the algorithm has unrestricted random-access to the underlying hypergraph), it is known that hypergraph sparsifiers require $\Omega(nr)$ bits to represent \cite{KKTY21a, KKTY21b}, and moreover, can be constructed in $\widetilde{O}(nr / \eps^2)$ bits of space. Thus, one interpretation of the work of \cite{KPS24d} is that hypergraph sparsifiers can be constructed in dynamic streams at the cost of only a $\log(m)$ times increase in the space complexity (ignoring $\polylog(n)$ factors), as compared to the static setting. However, this highlights two natural questions: 
\begin{enumerate}
    \item What is the space complexity of constructing hypergraph sparsifiers in insertion-only streams? Is the complexity essentially same as in the static setting, or can it be as large as the dynamic setting?
    \item More generally, how does the space complexity of hypergraph sparsification change as a function of the number of hyperedge deletions? Does it smoothly interpolate between the space complexity needed for insertion-only stream (no deletions) and the dynamic setting (unrestricted number of deletions)?
\end{enumerate}

\subsection{Our Contributions}

As our first contribution, we provide an answer to the first question above regarding the complexity of constructing sparsifiers in insertion-only streams:

\begin{theorem}\label{thm:insertionOnlyintro}
	There is an insertion-only streaming algorithm requiring $\widetilde{O}(n r / \eps^2)$ bits of space which creates a $(1 \pm \eps)$ cut-sparsifier for a hypergraph on $n$ vertices and hyperedges of arity $\leq r$, with probability $1 - 1 / \mathrm{poly}(n)$.
\end{theorem}

Specifically, this improves over the prior state of the art algorithms for constructing hypergraph sparsifiers in insertion-only streams by a factor of $\log(m)$ where $m$ denotes an upper bound on the number of hyperedges. This implies that (perhaps surprisingly) the complexity of the insertion-only setting \emph{mirrors} that of the static sparsification setting, and thus there is also a strict separation between the space complexity of the insertion-only and dynamic streaming settings, as the latter is known to require $\Omega(nr\log m)$ bits of space \cite{KPS24d}.

Because of this large separation, it is natural to ask if there is some other parameter which governs the space complexity of sparsifying hypergraphs in streams. 
As our second contribution, we show that this is indeed the case: if we parameterize the dynamic streaming setting by the maximum number of allowed hyperedge \emph{deletions}, then the space complexity smoothly interpolates between the insertion-only setting and the unrestricted dynamic setting:

\begin{theorem}\label{thm:boundedDeletionintro}
	For $k \geq 1$, there is a $k$-bounded deletion streaming algorithm requiring $\widetilde{O}(n r \log(k) / \eps^2)$\footnote{Technically, when $k = 1$, the term should be $\max(1, \log(k))$.} bits of space which creates a $(1 \pm \eps)$ cut-sparsifier for a hypergraph on $n$ vertices and hyperedges of arity $\leq r$, with probability $1 - 1 / \mathrm{poly}(n)$.
\end{theorem}

When $m$ is the maximum number of hyperedges in the stream, then the number of deletions is effectively bounded by $m$. Thus, the dynamic streaming setting is effectively the case when $k = m$, and the above theorem captures the space complexity in this setting. Likewise, as the number of deletions decreases and approaches $0$, the setting approaches the insertion-only setting, and the above theorem explains exactly the space savings that are achieved.

Finally, building off of the prior work of Jayaram and Woodruff \cite{JW18} in the bounded-deletion streaming model, we show that this space complexity is essentially \emph{optimal}:

\begin{theorem}\label{thm:lowerBoundIntro}
    Any streaming algorithm for $k$-bounded deletion streams, which for hypergraphs on $n$ vertices, of arity $r \leq n/2+1$, produces a $(1 \pm \eps)$ cut-sparsifier for $\eps < 1$, must use $\Omega(n r \log(k/n))$ bits of space.
\end{theorem}

In summary, this provides a complete picture of the space complexity of producing hypergraph sparsifiers in the streaming setting: as the number of deletions $k$ increases from $0$ to $m$, the space complexity grows by a factor of $\log(k)$ over the space complexity of the static sparsification regime, leading to a smooth phase transition in the space complexity of these algorithms. In the following subsection, we explain more of the techniques that go into these results.

\paragraph{Concurrent Work:} In concurrent work, Cohen-Addad, Woodruff, Xie, and Zhou \cite{CWXZ25} study hypergraph spectral sparsification (a strengthening of cut-sparsification) in the insertion-only regime and likewise show that the factor of $\log(m)$ can be shaved-off.

\subsection{Technical Overview}

\subsubsection{Importance Sampling for Hypergraph Cut Sparsification}\label{sec:importanceSampling}

To start, let us recap how we create hypergraph sparsifiers in the \emph{static} setting. After an extensive line of works studying hypergraph sparsification \cite{KK15, SY19, CKN20, KKTY21a, KKTY21b, JLS22, Lee23, KPS24}, the work of Quanrud \cite{Qua23} provided the simplest lens through which one can build hypergraph sparsifiers. Roughly speaking, given a hypergraph $H = (V, E)$, each hyperedge $e \in E$ is assigned a value $\lambda_e$ which is called its \emph{strength}. The strength of a hyperedge is intuitively a measure of the (un)importance of a hyperedge; the smaller the strength of the hyperedge $e$, this means $e$ is crossing smaller cuts in the hypergraph, and so we are more likely to need to keep $e$, while if the strength is larger, then there are many other hyperedges which cross the same cuts as $e$, and thus it is not as necessary for us to keep $e$. \cite{Qua23} showed that for a specific definition of strength, sampling each hyperedge (independently) at rate roughly $p_e \geq \log(n) / (\eps^2 \lambda_e)$ (ignoring constant factors), and assigning weight $1 / p_e$ to the surviving hyperedges then yields a $(1 \pm \eps)$ sparsifier with high probability. 

In fact, this procedure lends itself to a simple iterative algorithm for designing sparsifiers: starting with the original hypergraph $H$, we recover all of the hyperedges in $H$ whose strength is smaller than $\lambda \approx \log(n) / \eps^2$, and denote these hyperedges by $T^{(1)}$. Then, it must necessarily be the case that all hyperedges in $H - T^{(1)}$ have \emph{large strength}, and so we can afford to sample these hyperedges at rate $1/2$ (denote this sampled hypergraph by $H^{(1)}$). By the same analysis as above, it turns out that we can show that $T^{(1)} \cup 2 \cdot H^{(1)}$ will be a $(1 \pm \eps)$ sparsifier of $H$ with high probability. Now, it remains only to \emph{re-sparsify} $H^{(1)}$, which we can do by repeating the same procedure. Thus, after $\log(m)$ levels of this procedure (where $m$ is the starting number of hyperedges), we can recover a sparsifier of our original hypergraph. 

\subsubsection{Formal Definitions of Strength}

However, to continue our discussion, we will require the formal definition of strength, as well as some auxiliary facts about  strength in hypergraphs. The key notion that \cite{Qua23} introduces to measure \emph{strength} in hypergraphs is the notion of \emph{$k$-cuts} in hypergraphs (and here we adopt the language used by \cite{KPS24d}):

\begin{definition}\label{def:minkcut}
   For any $k \in [2..n]$, a $k$-cut in a hypergraph is defined by a $k$-partition of the vertices, say, $V_1, \dots V_k$. The \emph{un-normalized size of a $k$-cut} in an unweighted hypergraph is the number of hyperedges that are not completely contained in any single $V_i$ (we refer to these as the crossing hyperedges), denoted by $E[V_1, \dots V_k]$.

    The \emph{normalized size of a $k$-cut} in a hypergraph is its un-normalized size divided by $k-1$. We will often use $\minkcut(H)$ to denote the minimum normalized $k$-cut, defined formally as follows:
    \[
    \minkcut(H) = \min_{k \in [2..n]} \min_{V_1, \cup  \dots \cup V_k = V} \frac{|E[V_1, \dots V_k]|}{k-1}.
    \]
\end{definition}

Note that when we generically refer to a $k$-cut, this refers any choice of $k \in [2..n]$. That is, we are not restricting ourselves to a single choice of $k$, but instead allowing ourselves to range over any partition of the vertex set into any number of parts. 

The work of \cite{Qua23} established the following result regarding normalized and un-normalized $k$-cuts:

\begin{theorem}\cite{Qua23}
    Let $H$ be a hypergraph, with associated minimum normalized $k$-cut size $\minkcut(H)$. Then for any $t \in \Z^{+}$, and $k \in [2..n]$, there are at most $n^{O(t)}$ un-normalized $k$-cuts of size $\leq t \cdot \minkcut(H)$.
\end{theorem}

A direct consequence of the above is that in order to preserve all $k$-cuts (again, simultaneously for every $k \in [2, \dots n]$) in a hypergraph $H$ to a factor $(1 \pm \eps)$, it suffices to sample each hyperedge at rate $p \geq \frac{C \log(n)}{\eps^2 \minkcut(H)}$, and re-weight each sampled hyperedge by $1/p$. 

Similar to Bencz\'ur and Karger's \cite{BK96} approach for creating $\widetilde{O}(n / \eps^2)$ size graph sparsifiers, Quanrud \cite{Qua23} next uses this notion to define \emph{$k$-cut strengths} for each hyperedge. To do this, we fix a minimum normalized $k$-cut, with $V_1, V_2, ..., V_k$ denoting the partition of the vertices created by this cut. For any hyperedge crossing this minimum normalized $k$-cut, we define its strength to be exactly $\minkcut(H)$. For the remaining hyperedges (i.e, those which are completely contained within the components $V_1, \dots V_k$), their strengths are determined recursively (within their respective induced subgraphs) using the same scheme. This allows Quanrud \cite{Qua23} to calculate sampling rates of hyperedges, which when sampled, approximately preserve the size of every $k$-cut (for all $k \in [2, n]$). Note that just as in the graph setting, the reciprocal sum of strengths is bounded, which allows for convenient bounds on the number of low strength hyperedges: 

\begin{claim}\cite{Qua23}\label{clm:lowstrengthBound}
    Let $H = (V, E)$ be a hypergraph on $n$ vertices. Then,
    \[
    \sum_{e \in E} \frac{1}{\lambda_e} = n-1.
    \]
\end{claim}

However, the power of the strength definition extends beyond just identifying sampling rates of hyperedges, and can also be used to identify \emph{sets of vertices} which can be contracted away. In particular, we can define the notion of the \emph{strength of a component}, as we do below:

\begin{definition}
    For a subset of vertices $ S \subseteq V$, we say that the \emph{strength of $S$ in $H$} is $\lambda_S = \min_{e \in H[S]} \lambda_e$. That is, when we look at the induced subgraph from looking at $S$, $\lambda_S$ is the minimum strength of any edge in this induced subgraph. 
\end{definition}

We will take advantage of the following fact when working with these \say{contracted} versions of hypergraphs:

\begin{claim}\label{clm:minkcutcontracted}[\cite{KPS24d}]
    Let $H$ be a hypergraph, and let $V_1, \dots V_k$ be a set of connected components of strength $> \kappa$. Then, the hyperedges of strength $\leq \kappa$ in $H$ are exactly those hyperedges of strength $\leq \kappa$ in $H/(V_1, \dots V_k)$, where we use $H/(V_1, \dots V_k)$ to denote the hypergraph where $V_1, \dots V_k$ have each been \emph{contracted} to their own super-vertices.
\end{claim}

This final claim will be very important for our algorithm. In particular, it implies that once certain components have become sufficiently strongly connected, we no longer have to worry about recovering low-strength hyperedges \emph{within} the components, and can instead focus only on recovering hyperedges that \emph{cross between} such components. However, before diving more into the details of this approach, we provide a more detailed re-cap of how prior work \cite{KPS24d} recovers low-strength hyperedges generically.

\subsubsection{The Work of \cite{KPS24d}}

With this formal notion of strength now in hand, we can formally present the algorithm discussed in the previous subsection for sparsification (essentially the framework of \cite{BK96, AGM12, GMT15, KPS24d}):

\begin{algorithm}[H]
\caption{SimpleSparsification($H, \eps$)}\label{alg:introSimple}
Let $H_0 = H$, let $C$ be a sufficiently large constant.  \\
\For{$i = 0, 1, \dots \log(m)$}{
Let $F_i$ be all hyperedges in $H_i$ of strength $\leq 2C \log(n) / \eps^2$. \\
Store $F_i$. \\
Let $H_{i+1}$ be hyperedges in $(H_i - F_i)$ sampled at rate $1/2$.
}
\Return{$\cup_i 2^i \cdot F_i$.}

\end{algorithm}

This approach is exactly what was used by \cite{KPS24d} when designing their hypergraph sparsifiers for dynamic streams. 

Indeed, their primary contribution was a \emph{linear sketch} which can be used to exactly recover these low-strength hyperedges at each level of the algorithm. Recall that a linear sketch is simply a set of linear measurements of the hypergraph $H$ and thus is \emph{directly implementable} as a dynamic streaming algorithm on hypergraphs, as both insertions and deletions can be modeled as linear updates. Formally, when a hypergraph on $n$ vertices is viewed as a vector in $\{0,1\}^{2^n}$, then a linear measurement of size $s$ is obtained by mutliplying a (possibly random) $s \times 2^n$ matrix with this vector. Inserting a hyperedge is simply a $+1$ update in the coordinate corresponding to the hyperedge, and a deletion is simply a $-1$.

With this established, we can now summarize the space complexity resulting from the linear sketching implementation of \cite{KPS24d}:
\begin{enumerate}
    \item At each of the $\log(m)$ levels of sampling in the above algorithm, the linear sketch \cref{sec:importanceSampling} can be used to recover the low strength edges. This is accomplished by storing $\log(m)$ independent copies of a sketch for recovering low-strength hyperedges (one copy at each level).
    \item Within each individual sketch (at a fixed level of the sampling process), there is a specific linear sketch stored for the neighborhood of the $n$ vertices. 
    \item Each such vertex neighborhood sketch requires space $\widetilde{O}(r  / \eps^2)$ bits.
\end{enumerate}
In total then, this yields a linear sketch (and hence a dynamic streaming algorithm) which stores $\widetilde{O}(\log(m) \cdot n \cdot r / \eps^2)$ bits. Once this sketch has been stored, the algorithm can simply iteratively recover the low strength edges at each of the $\log(m)$ levels of sampling, thereby recovering a sparsifier of the original hypergraph. 

\subsubsection{Optimizing the Complexity in Insertion-only Streams}

At first glance, it may seem that none of these parameters from the work of \cite{KPS24d} can be optimized in the insertion-only setting: indeed, there are $m$ hyperedges in the hypergraph initially, and thus any iterative procedure will require $\Omega(\log(m))$ levels before exhausting the hypergraph if the sampling rate is $1/2$. Likewise, the $n$ vertices are fixed, and the linear sketch designed by \cite{KPS24d} requires storing the linear sketches for each vertex. Lastly, the complexity of the sketches for each vertex cannot hope to be improved, as simply recovering a single hyperedge yields $\Omega(r)$ bits of information. Thus, it may seem that one cannot build on top of this framework while achieving a space complexity that beats $O(n r \log(m))$ bits of space.

However, our first theorem shows that by cleverly merging and contracting vertices as hyperedges are inserted, we can in fact improve the space complexity. Perhaps counterintuitively, our optimization actually comes from \emph{decreasing the number of vertices} (the parameter $n$ above). 

To illustrate, let us consider a sequence of hyperedge insertions, and let us suppose that at some point in time, a large polynomial number of hyperedges have been inserted (say, some $n^{1002}$ hyperedges). As one might expect, if so many hyperedges have been inserted, there will naturally emerge certain components in the hypergraph which are very strongly connected. More formally, if we revisit \cref{clm:lowstrengthBound}, we can observe that the number of hyperedges of strength $< n^{1000}$, \emph{must be} less than $n^{1001}$. This implies that among the $n^{1002}$ hyperedges which have been inserted, the \emph{vast majority} are high strength hyperedges, and thus also define many high strength components. 

As a consequence, after these hyperedges are inserted, there will be components $C \subseteq V$ for which all of the hyperedges in the component $C$ are \emph{high-strength hyperedges}, and therefore do not need to be recovered in order to perform sampling by \cref{clm:minkcutcontracted}. Algorithmically, because we are dealing with an \emph{insertion-only} stream, once such a component $C$ becomes a high-strength component, it will forever remain a high-strength component, and thus, as per \cref{clm:minkcutcontracted}, we can effectively contract this component away. 

Thus, our algorithmic plan is principled: we will estimate the strength of components as hyperedges are inserted (separately for each level of the $\log(m)$ levels of sampling in the hypergraph), and whenever a component gets sufficiently large strength, we \emph{irrevocably contract} the component away to a single super-vertex. We do this for the hypergraphs at each of the $\log(m)$ levels of sampling. Thus, there are two key points that must be shown:
\begin{enumerate}
    \item We must show that contracting these vertices saves space in our sketch.
    \item We must show that we can estimate the strength in $\widetilde{O}(nr)$ bits of space in the (insertion-only) streaming setting. 
\end{enumerate}

In what follows, we explain how we achieve both of these goals.

\subsubsection{Saving Space by Contracting Vertices}

First, we show that we can save space by contracting vertices. Let us consider the top level of sampling in the hypergraph. As mentioned above, as hyperedges are inserted, there will naturally become certain strongly connected components.
So, let us denote one such component by $C$. Now, once this component is strong, it remains strong, and so we can be sure that we do not need to recover any hyperedges \emph{within} the component, as per \cref{clm:minkcutcontracted}. So, instead of storing the individual linear sketches $\mathcal{S}_v$ for the vertices $v \in C$, we instead \emph{add the sketches together}, yielding $\mathcal{S}_C = \sum_{v \in C} \mathcal{S}_v$. Note that this operation is not reversible, and we are in fact \emph{losing information} about the hypergraph when we perform this addition. However, this is the same reason that we will save some space: indeed, if there are $k$ strong components, we end up storing only the linear sketches of \cite{KPS24d} for the $k$ super-vertices, leading to a total space usage of $\widetilde{O}(k r / \eps^2)$ bits in a single level, as opposed to the $\widetilde{O}(n r / \eps^2)$ bits in \cite{KPS24d} (note there is no $\log(m)$ here as we are only looking at the top level of sampling for now). 

Unfortunately, this analysis alone is not sufficient for us to claim any space savings. Indeed, it is possible that (for instance) in the top level of sampling, there are \emph{no} strong components. It follows then that we cannot add together any linear sketches, and must instead pay $\widetilde{O}(n r / \eps^2)$ at this level of sampling. However, this is where we now use a key bound on the number of low-strength edges \cref{clm:lowstrengthBound} \cite{Qua23}. If there are no components of strength say, greater than $n^{1000}$, \emph{then there must be fewer than $n^{1001}$ hyperedges remaining}. It follows then that instead of storing this sketch of size $\widetilde{O}(n r / \eps^2)$ bits at each of the $\log(m)$ levels of sampling, we must only store it at $\log(n^{1001}) = O(\log(n))$ levels of sampling, thereby replacing this $\log(m)$ factor with a $\log(n)$ factor.

This argument as we have presented it though is far from general and must be extrapolated to work on all instances. In particular,  it is possible that at different levels of sampling, the strong components are different, and thus there is no single set of strong components that we can look at. To address this, we make the observation that the strong components form a laminar family. That is to say, the strong components at the $i+1$st level of sampling are a refinement of the strong components at the $i$th level of sampling. Thus, across all $\log(m)$ levels of sampling, the number of distinct strong components that appear is bounded by $O(n)$. Likewise, because any component of strength $\geq n^{1000}$ is merged away, by the same logic as above, each strong component must have $\leq n^{1001}$ incident hyperedges (i.e., hyperedges which touch this component, as well as some other component). Thus, each component that appears will only have a non-empty neighborhood for $O(\log(n))$ levels of sampling (after which point we do not need to store anything - as the sketch of an empty neighborhood is empty). 

In summary then, for each of the $O(n)$ strong component that appears, we store the sketch from \cite{KPS24d} of size $\widetilde{O}(r / \eps^2)$ for $O(\log(n))$ different levels of sampling. This then yields the desired complexity of $\widetilde{O}(n r /\eps^2)$ bits of space for the final algorithm. 

\subsubsection{Identifying Strong Components}

Finally, we show that we can approximately find the strong components in the hypergraph in an insertion-only stream. Fortunately, the foundation for this algorithm was presented in \cite{KPS24d}: let us consider again the hypergraph at the top level of sampling, which we denote by $H$. Simultaneously, we consider an auxiliary hypergraph $\hat{H}$ which is the result of sampling the hyperedges of $H$ at rate $1 / n^{1000}$. 

As shown in \cite{KPS24d}, it turns out the connected components in $\hat{H}$ exactly correspond to the strong components in $H$. Thus, in the insertion-only streaming model, this admits an exceedingly simple implementation: as hyperedges arrive,  we sample them at rate $1 / n^{1000}$, and keep track of the components. Then, whenever a new component is formed, we simply add together the corresponding linear sketches as discussed in the previous section (i.e., merging those vertices together). Note that storing the set of connected components can be done in $\widetilde{O}(n)$ bits of space, and thus across $\log(m)$ levels of sampling, requires only $\widetilde{O}(n \log(m))$ bits. Because $m \leq n^r$, this then yields the desired space complexity. These ideas then suffice for the insertion-only implementation.

\subsubsection{Generalizing to Bounded-Deletions and Remarks}

The key observation for generalizing the bounded deletion setting is that once a component has strength $k + n^{1000}$, then after any sequence of $k$ deletions, the strength of the component remains \emph{at least} $n^{1000}$. Thus, instead of adding samplers together after the components reach strength $n^{1000}$, we instead add samplers together once the strength reaches $k + n^{1000}$. Observe then that the $\log(k)$ term is a natural artifact: each strong component can have a non-empty neighborhood of incident hyperedges for $\log(k+n)$ levels of sampling, as opposed to simply $O(\log(n))$ levels of sampling, and this yields the final complexity. 

We now finish with some remarks:

\begin{enumerate}
    \item The primary work-horse of \cite{KPS24d} is a linear sketch known as an $\ell_0$-sampler, and it is tempting to simply try to replace the $\ell_0$-samplers via linear sketching in their paper with an $\ell_0$-sampler specifically for insertion-only streams, thereby saving space. However, such a replacement would necessitate an entirely new analysis: even though the stream itself may not have deletions, the process of recovering hyperedges, merging vertices together, and more \emph{all rely} on the ability to linearly add together $\ell_0$-samplers (which causes deletions). While the aforementioned approach may work, it would not build on the existing framework. The same goes for the bounded deletion setting, where it is tempting to use $\ell_0$-samplers defined for bounded-deletion streams (as discussed in \cite{JW18}).
    \item There are several subtleties that arise in the analysis regarding the estimation of strong components, as this will never be an \emph{exact} decomposition of the graph. We instead provide upper and lower bounds on the strength of the components and remaining hyperedges, and use this to facilitate our analysis. We defer a more complete description to the technical sections below. 
    \item Likewise, in the bounded-deletion setting, there is considerable difficulty in optimizing the dependence on $k$ to \emph{not be} $\log^2(k)$. Roughly speaking, this is because the $k$ dependence tries to show up in both (1) the number of levels of sampling in which a strong component has a non-empty neighborhood (a $\log(k)$ factor as described above) and (2) the support size of the $\ell_0$-samplers that are needed when using the sketch of \cite{KPS24d} (another factor of $\log(k)$). We use a more refined analysis along with a \emph{second} round of component merging to bypass this other factor of $\log(k)$.
    \item The lower bound follows from the augmented index problem along with an argument from \cite{JW18} on bounding the complexity of this problem in the bounded-deletion setting. We omit a full discussion here, as the analysis presented in \cref{sec:lowerbound} is concise.
\end{enumerate}

\subsection{Organization}

In \cref{sec:prelim} we introduce more formal definitions of our model and statements of prior work. In \cref{sec:insertion}, we prove \cref{thm:insertionOnlyintro}, in \cref{sec:booundedDeletion} we prove \cref{thm:boundedDeletionintro}, and finally in \cref{sec:lowerbound}, we present \cref{thm:lowerBoundIntro}.

\section{Preliminaries}\label{sec:prelim}

\subsection{Definitions}

First, we recap the definition of a hypergraph, and the notion of a hypergraph cut-sparsifier.

\begin{definition}
	A hypergraph $H = (V,E)$ is given by a vertex set $V$ of size $n$, and a set of hyperedges $E$ where each $e \in E$ is an arbitrary subset of $V$. For a hyperedge $e$, we say that the \emph{arity} of the hyperedge is $|e|$. The arity of the hypergraph is $\max_{e \in E} |e|$. The \emph{support size} of the hypergraph is the number of distinct hyperedges, which is denoted by $m$.
	
	Occasionally, a hypergraph may be given by a triple $(V, E, w)$ where $w$ is a weight function $E: \mathbb{R}^{\geq0}$. When a weight function is omitted, the implication is that every hyperedge has weight $1$. 
\end{definition}

Given a hypergraph, we can also define the notion of \emph{cuts}:

\begin{definition}
	For a hypergraph $H =  (V, E, w)$ and a subset $S \subseteq V$, we say that the hyperedges cut by $S$ are 
	\[
	\mathrm{cut}_H(S) = \{ e \in E: e \cap S \neq \emptyset \wedge 	e \cap \bar{S} \neq \emptyset\}.
	\]
	The weight of the cut is then given by 
	\[
	|\mathrm{cut}_H(S)| = \sum_{e \in \cut_H(S)} w_e. 
	\]
	\end{definition}
	
We are now ready to define a cut-sparsifier:

\begin{definition}
	For a hypergraph $H$, and a parameter $\eps \in (0,1)$, a $(1 \pm \eps)$ cut-sparsifier of $H$ is a hypergraph $\widetilde{H}$ such that (simultaneously) for every $S \subseteq V$:
	\[
		|\mathrm{cut}_{\widetilde{H}}(S)| \in (1 \pm \eps)	|\mathrm{cut}_{H}(S)|.
	\]
\end{definition}

In this work, we will be particularly interested in the streams of hyperedges:

\begin{definition}
	A stream of hyperedges is given by a sequence $(e_1, \delta_1), (e_2, \delta_2), \dots (e_i, \delta_i), \dots$, where each $e \subseteq V$, and each $\delta_i \in \{\pm 1\}$. A pair $(e, -1)$ corresponds to a hyperedge being deleted, and a pair $(e, 1)$ corresponds to a hyperedge being inserted. The number of deletions in the stream is given by the number of tuples where $\delta_i = -1$.
	
	A stream is a $k$-deletion stream if there are at most $k$ deletions occurring in the stream. If $k = 0$, we refer to this as an insertion-only stream.
	\end{definition}
	
One useful tool for working with streams is the notion of a linear sketch:

\begin{definition}
	A linear sketch of a hypergraph of arity $r$ is specified by a (possibly randomized) sketching matrix $M \in \mathbb{Z}^{s \times \binom{n}{\leq r}}$. For a hypergraph $H$, the sketch is given by representing $H$ as a vector in $\mathbf{1}_H \in \zo^{\binom{n}{\leq r}}$ such that $(\mathbf{1}_H)_e = \mathbf{1}[e \in H]$, and then multiplying $M \mathbf{1}_H$. 
	
	The number of entries in the sketch is then $s$, though the number of bits required to represent the sketch may be larger (since the values can be arbitrary integers).
\end{definition}

\subsection{Useful Notions from Prior Work}

Now, we recap some useful results from prior work. In particular, we start with the definition of \emph{strength} in hypergraphs:

\begin{definition}
	For a hypergraph $H = (V, E)$, the \emph{minimum normalized $k$-cut} is defined to be 
	\[
	\min_{k \in [n]} \min_{V_1 \cup V_2 \cup \dots \cup V_k = V} \frac{|E[V_1, \dots V_k]|}{k-1}.
	\]
	$|E[V_1, \dots V_k]|$ refers to the number of edges which cross between (any subset) of $V_1, \dots V_k$. This is a generalization of the notion of a $2$-cut in a graph, which is traditionally used to create cut sparsifiers in ordinary graphs. Further, note that $V_1, \dots V_k$ form a partition of $V$. As mentioned in the introduction, we will often use the following to denote the minimum normalized $k$-cut:
	\[
	\Phi(H) = \min_{k \in [n]} \min_{V_1, \cup  \dots \cup V_k = V} \frac{|E[V_1, \dots V_k]|}{k-1}.
	\]
	We also refer later to \emph{un-normalized $k$-cuts}, which is simply $|E[V_1, \dots V_k]|$, for some partition $V_1, \dots V_k$ of $V$.
\end{definition}

Now, to define strength, we iteratively use the notion of the minimum $k$-cut.
\begin{definition}
	Given a hypergraph $H = (V, E)$, let $\Phi(H)$ be the value of the minimum normalized $k$-cut, and let $V_1, \dots V_k$ be the components achieving this minimum. For every edge $e \in E[V_1, \dots V_k]$, we say that $\lambda_e = \Phi(H)$. Now, note that every remaining edge is contained entirely in one of $V_1, \dots V_k$. For these remaining edges, we define their strength to be the strength inside of their respective component.
\end{definition}

\begin{remark}\label{rmk:monotone}
	Note that the strengths assigned via the preceding definition are non-decreasing. Indeed if the minimum normalized $k$-cut has value $\phi$ and splits a graph into components $V_1, \dots V_k$, it must be the case that the minimum normalized $k$-cuts in each $H[V_i]$ are $\geq \phi$, as otherwise one could create an even smaller original normalized $k$-cut by further splitting the component $V_i$. See \cite{Qua23, KPS24d} for a longer discussion. 
\end{remark}

We will also refer to the strength of a component.

\begin{definition}
	For a subset of vertices $ S \subseteq V$, we say that the \emph{strength of $S$ in $H$} is $\lambda_S = \min_{e \in H[S]} \lambda_e$. That is, when we look at the induced subgraph from looking at $S$, $\lambda_S$ is the minimum strength of any edge in this induced subgraph. 
\end{definition}

\begin{definition}\label{def:contractedHypergraph}
	For a hypergraph $H$ and partition $V_1, \dots V_k$ of the vertex set, let $H / (V_1, \dots V_k)$ denote the hypergraph obtained by contracting all vertices in each $V_i$ to a single vertex. For a hyperedge $e \in H$, we say that the corresponding version of $e \in H / (V_1, \dots V_k)$ (denoted by $e / (V_1, \dots V_k)$) is incident on a super-vertex corresponding to $V_i$ if there exists $v \in V_i$ such that $v \in e$.
\end{definition}

We will take advantage of the following fact when working with these \say{contracted} versions of hypergraphs:

\begin{claim}
	Let $H$ be a hypergraph, and let $V_1, \dots V_k$ be a set of connected components of strength $> \kappa$. Then, the hyperedges of strength $\leq \kappa$ in $H$ are exactly those hyperedges of strength $\leq \kappa$ in $H/(V_1, \dots V_k)$.
\end{claim}

Now, we are ready to summarize some important results from \cite{KPS24d}:

\begin{theorem}\label{thm:kpssketch}[\cite{KPS24d}]
	There is a linear sketch of size $\widetilde{O}(n r \kappa \log(m) )$ 
		bits, which for arbitrary  hypergraphs $H$ on $n$ vertices, $\leq m$ hyperedges, of arity $\leq r$, recovers exactly the hyperedes of $H$ with strength $\leq \kappa$. The linear sketch is composed of $n$ separate linear sketches of size $\widetilde{O}(r \kappa \log(m) )$ for each vertex $v \in V$.
\end{theorem}

An important building block in the above theorem is the following linear sketch:

\begin{theorem}\label{thm:gmtsketch}[\cite{GMT15}]
	There is a linear sketch of size $\widetilde{O}(n r  \log(m) )$ 
	bits, which for arbitrary hypergraphs $H$ on $n$ vertices, $\leq m$ hyperedges, of arity $\leq r$, recovers exactly the connected components of $H$ with high probability. The linear sketch is composed of $\log(n)$ $\ell_0$ samplers initialized for each vertex $v \in V$. 
\end{theorem}

The following algorithm has appeared in many sparsification settings, and yields a simple procedure for sparsifying hypergraphs:

\begin{algorithm}
	\caption{SparsifyHypergraph$(H, \eps)$}\label{alg:simpleSparsify}
	Let $i = 0$, and for all $j: H_j = H$. \\
	\While{$H_i$ is not empty}{
	Let $F_i$ be all hyperedges in $H_i$ of strength $\leq C \log(n) / \eps^2$, for a large constant $C$. \\
	Let $H_{\text{intermediate}} = H_i - F_i$. \\
	Subsample the hyperedges of $H_{\text{intermediate}}$ at rate $1/2$. \\
	Set $H_{i+1} = H_{\text{intermediate}}$. \\
	$i \leftarrow i+1$. 
	}
	\Return{$\widetilde{H} = \bigcup_{i} 2^i \cdot F_i$.}
\end{algorithm}

A basic analysis shows that the above algorithm yields a $(1 \pm O(\eps \cdot \log(m)))$-sparsifier of $H$ with probability $1 - 1 / \mathrm{poly}(n)$. This is essentially due to composing $\log(m)$ levels of sparsifiers, and naively combining their error. Thus, in order to get a $(1 \pm \eps)$ sparsifier, one would need to instead invoke the algorithm with a parameter $\eps' = \eps / \log(m)$, which leads to extra factors of $\log(m)$ in the sparsifier size (as one must recover more hyperedges in each level).

The work of \cite{KPS24d} showed the following general fact about error-accumulation when sparsifying hypergraphs:

\begin{theorem}\label{thm:kpsaccuracy}[\cite{KPS24d}]
	For any choice of $\eps' \leq \eps / \log^2(n / \eps)$, SparsifyHypergraph$(H, \eps')$ returns a $(1 \pm \eps)$ hypergraph cut-sparsifier to $H$ with probability $1 - 1 / \mathrm{poly}(n)$. 
	\end{theorem}
	
Now, we recap a useful auxiliary algorithm that we will use:

\begin{algorithm}[H]
	\caption{FindStrongComponents$(H)$}\label{alg:findStrongComp}
	$\widetilde{H}^{(0)} = H$. \\
	\For{$i = 1, \dots r \log(n)$}
	{
		Let $\mathrm{CC}^{(i)}$ be the connected component structure of $\widetilde{H}^{(i)}$. \\
		Let $\widetilde{H}^{(i+1)} = \widetilde{H}^{(i)}$ with edges sampled at rate $1/2$. \\
	}
	\Return{$\{\mathrm{CC}^{(i)}: i \in [r \log(n)] \}$}
\end{algorithm}

We have the following claims from \cite{KPS24d}:

\begin{claim}\label{clm:subsamplingGivesStrong}
	Let $H$ be a hypergraph, and let $i \in [r \log(n)]$. Let $H_i$ be hypergraph resulting from \cref{alg:simpleSparsify} at the $i$th level of sampling. Then, the components $\mathrm{CC}^{(i + 20 \log(n))}$ from \cref{alg:findStrongComp} are components of strength $\geq n^{15}$ in  $H_i$ with probability $1 - 2^{-\Omega(n)}$.
\end{claim}

\begin{claim}
	Let $H$ be a hypergraph, and let $i \in [r \log(n)]$. Let $H_i$ be hypergraph resulting from \cref{alg:simpleSparsify} at the $i$th level of sampling. Then, any hyperedge of strength $\geq n^{100}$ in $H_i$ is completely contained in some component $S \in \mathrm{CC}^{(i + 20 \log(n))}$ from \cref{alg:findStrongComp} with probability $1 - 2^{-\Omega(n)}$.
\end{claim}

As a corollary, we have that if we merge each component in  $\mathrm{CC}^{(i + 20 \log(n))}$ into a super-vertex in $H_i$, then the number of crossing hyperedges in $H_i$ is bounded. 

\begin{corollary}\label{cor:boundedHyperedges}
	Let $H$ be a hypergraph, and let $i \in [r \log(n)]$. Let $H_i$ be hypergraph resulting from \cref{alg:simpleSparsify} at the $i$th level of sampling. Then, the hypergraph $H_i /  \mathrm{CC}^{(i + 20 \log(n))}$ has $\leq n^{101}$ hyperedges, and any hyperedge of strength $\leq n^{15}$ in $H_i$ is still in $H_i /  \mathrm{CC}^{(i + 20 \log(n))}$.
\end{corollary}

\begin{proof}
	Every hyperedge of strength $\geq n^{100}$ has been contracted away. Hence, the only remaining hyperedges are those of strength $\leq n^{100}$ in $H_i$, of which there can be at most $n^{101}$.
	
	To see the second claim, each component $\mathrm{CC}^{(i + 20 \log(n))}$ in $H_i$ has strength $\geq n^{15}$. Hence, any hyperedges of strength $\leq n^{15}$ are still crossing between these components. 
\end{proof}

This final corollary is what allows \cite{KPS24d} to essentially replace the $\log(m)$ dependence in their linear sketch with a $\log(n)$ factor, as the linear sketch is only ever opened on the \emph{contracted} version of the hypergraph, where there are now fewer hyperedges. However, because they carefully contracted, they are also guaranteed that the low-strength edges in the contracted version of the hypergraph are \emph{the same} as the low-strength edges in the original version. Thus, their linear sketch succeeds at recovering from the original hypergraph, despite working on a hypergraph with many fewer hyperedges.

However, as shown by \cite{KPS24d}, this approach can only ever get down to $nr \log(m)$ bits of space (as indeed this is optimal for a linear sketch).
In this work, we instead want to get rid of this final $\log(m)$ factor when we move to the \emph{insertion-only} regime. With this, we now have the necessary context for deriving our results. We delve into this in the following section.
	
\section{Building Sparsifiers in Insertion-Only Streams}\label{sec:insertion}

In this section, we will show the following:

\begin{theorem}\label{thm:insertionOnly}
	There is an insertion-only streaming algorithm requiring $\widetilde{O}(n r / \eps^2)$ bits of space which creates a $(1 \pm \eps)$ cut-sparsifier for a hypergraph on $n$ vertices and hyperedges of arity $\leq r$, with probabiltiy $1 - 1 / \mathrm{poly}(n)$.
\end{theorem}

\begin{remark}
	Observe that the above theorem has \emph{no} dependence on the length of the stream. In particular the number of hyperedges $m$ can be potentially as large as $\binom{n}{\leq r}$. Nevertheless, the above algorithm always uses $\widetilde{O}(n r / \eps^2)$ bits of space, which nearly-matches the best possible space used in a static sparsifier. 
\end{remark}

\subsection{Insertion-Only Improvement}

With the \say{strong component} contractions introduced in the previous section, this effectively reduces the support of the linear sketch used to recover low-strength hyperedges to $\mathrm{poly}(n)$, meaning that we are storing only $\widetilde{O}(n r / \eps^2)$ bits for each of the low-strength recovery sketches. However, there is one final optimization that we can perform in the insertion-only regime. 

First, let us recall how exactly the linear sketch for recovering low-strength hyperedges from \cite{KPS24d} works: for each of the $n$ vertices, the linear sketch stores $\ell_0$ samplers for the neighborhood of the vertex (under different rates of \say{fingerprinting}). To get the $\ell_0$ samplers for an arbitrary contracted component, one must only \emph{add together} the $\ell_0$ samplers of the constituent vertices for that component. However, there is one key limitation in the dynamic streaming (linear sketching) setting. Although we may have information about the strong components at some time step of the algorithm, we must still maintain individual samplers for each vertex. Because the stream can \emph{remove} already inserted hyperedges, it is possible that what was once a strong connected component $S \subseteq V$ at some timestep $t$ is no longer a strong connected component at some timestep $t'$. Once the component is no longer strong, we must store $\ell_0$ samplers for the individual vertices it contains. This forces us to store $\Omega(n)$ $\ell_0$ samplers at each level of the sparsification algorithm, hence leading to the $\Omega(nr\log(m))$ bit complexity lower-bound. 

However, in the insertion-only streaming model we can actually bypass this issue. Indeed, once a hyperedge is inserted, it will never be deleted. In particular, this means that as the algorithm progresses and more hyperedges are inserted, our estimates of strong components $\mathrm{CC}^{(i)}$ is \say{monotonically} contracting. That is, our strong components are only ever being merged together, never being split apart. Thus, once we identify a strong component at some level of the algorithm, we can be sure that we do not need to store individual $\ell_0$ samplers for the constituent vertices, and can instead simply store a single $\ell_0$ sampler for the entire component as a whole. We detail this new streaming algorithm below:

\begin{algorithm}[H]
	\caption{StreamingAlgorithm($H, \eps)$)}\label{alg:buildingSketch}
	To start, initialize $\mathrm{CC}^{(i)} = [n]$ for each $i$. \\
	Initialize the sketch of \cref{thm:kpssketch} for each vertex $v \in V$, for each $i \in [r \log(n)]$ using $\kappa  = \frac{C \log^5(n / \eps)}{ \eps^2}$. Denote the sketch for vertex $v$, level $i$, by $\mathcal{S}_{v,i}$. \\
	\For{$(e, 1)$ arriving in the stream}{
	\For{$i \in \{0, 1, \dots  r \log(n)\}$}{
	Add $e$ to every linear sketch $\mathcal{S}_{C, i}: C \in \mathrm{CC}^{(i)}$. \\
	Update $\mathrm{CC}^{(i)}$ to merge any components connected by $e$. \\
	If any components $C_1, \dots C_{\ell}$ are merged together in $\mathrm{CC}^{(i)}$  to create a larger strong component $C' = C_1 \cup C_2 \cup \dots C_{\ell}$, set $\mathcal{S}_{C', i - 20 \log(n)} = \mathcal{S}_{C_1, i - 20 \log(n)} + \mathcal{S}_{C_2, i - 20 \log(n)}  + \dots + \mathcal{S}_{C_{\ell}, i - 20 \log(n)} $, and delete $\mathcal{S}_{C_1, i - 20 \log(n)}, \dots \mathcal{S}_{C_{\ell}, i - 20 \log(n)}$. \\
	With probability $1/2$; end \textbf{for}, otherwise, continue.
	}
	}
\end{algorithm}

Note that once we have built the sketch, we can also use it to easily recover a sparsifier:

\begin{algorithm}[H]
		\caption{RecoverSparsifier($H, \eps)$)}\label{alg:recoverSparsifier}
		\For{$i \in \{0, 1, \dots  r \log(n)\}$}{
			Open the linear sketches $\mathcal{S}_{C, i}: C \in \mathrm{CC}^{(i)}$ in accordance with \cref{thm:kpssketch}. \\
			Denote the recovered hyperedges by $F_i$. \\
			Remove these recovered hyperedges from sketches at levels $j > i$. 
		}
		\Return{$\bigcup_i 2^i \cdot F_i$.}
\end{algorithm}

In the next section, we analyze the above algorithm (both space and accuracy).

\subsection{Analysis of \cref{alg:buildingSketch} and \cref{alg:recoverSparsifier}}

First, we focus on the space usage of \cref{alg:buildingSketch}. We have the following claim:

\begin{claim}
	Fix a time step $t$ of the stream, and let $\mathrm{CC}^{(i)}: i \in [r \log(n)]$ denote the estimates of the strong connected components (across all levels of sampling). Then,
	\[
	|\{S \subseteq V: \exists i \in [r \log(n)]: S \in \mathrm{CC}^{(i)} \}| \leq O(n).
	\]
\end{claim}

\begin{proof}
	This follows because the connected components we recover form a laminar family. I.e., the connected components in $\mathrm{CC}^{(i)}$ are a refinement of the connected components in $\mathrm{CC}^{(i-1)}$.
\end{proof}

Note that this claim is not enough to give us anything substantial as is. Indeed, for each component we will store $\mathrm{polylog}(n) / \eps^2$ $\ell_0$-samplers, across each of the levels of sampling in which the component appears. In the worst case, a component will be present for $\Omega(\log(m))$ levels of sampling, and therefore require $\Omega(\log(m) \polylog(n) / \eps^2)$ $\ell_0$-samplers. Across the $O(n)$ components which appear, this would not offer us any substantial savings in space. 

Our key observation is actually in the fact that the components we see \emph{can only have a non-zero neighborhood for $O(\log(n))$ rounds of sampling}. Thus, even though we may store many $\ell_0$ samplers for a given component, \emph{most} of these $\ell_0$-samplers are empty, and can therefore be stored with a single bit. We formalize this claim below:

\begin{claim}\label{clm:numberLevelsInsertion}
	Fix a timestep $t$. Let $S$ be a component which appears in $\mathrm{CC}^{(i)}$ for some $i \in [r \log(n)]$. Then, there are only $O(\log(n))$ levels of sampling in \cref{alg:buildingSketch} where the neighborhood of $S$ is non-empty with probability $1 - 1 / n^{99}$.
\end{claim}

\begin{proof}
	Let $j$ denote the first round in which $S$ appears in $\mathrm{CC}^{(j)}$. Then, by \cref{cor:boundedHyperedges}, the number of hyperedges total in $H_{j - 20 \log(n)} / \mathrm{CC}^{(j)}$ is at most $n^{101}$ (here, we use $H_{j - 20\log(n)}$ to denote the hypergraph after $j - 20\log(n)$ levels of sampling in the above algorithm). In particular, this means that the number of hyperedges which can be incident on $S$ in $H_{j - 20 \log(n)} / \mathrm{CC}^{(j)}$ is also bounded by $n^{101}$. Thus, with high probability, after say, $200 \log(n)$ levels of sampling, all hyperedges incident on $S$ will have been removed, and hence its neighborhood is empty.
\end{proof}

Now, if the neighborhood of $S$ is empty, then to store $\ell_0$ samplers for this neighborhood is entirely trivial. Using this, we can bound the total space required by \cref{alg:buildingSketch}.

\begin{claim}\label{clm:sketchSpaceInsertion}
	At any timestep $t$, the total space required by \cref{alg:buildingSketch} is $\widetilde{O}(nr / \eps^2)$ bits with probability $1 - 1 / \mathrm{poly}(n)$.
\end{claim}

\begin{proof}
	First, for at most $\log(m) \leq r \log(n)$ levels, we store $1$ bit for $\leq n$ vertices' empty $\ell_0$ samplers. Thus, the total contribution from these empty samplers is $\widetilde{O}(nr)$ bits. 
	
	Now, for the non-empty $\ell_0$ samplers, we must only store these for $O(n)$ strong connected components, for $O(\log(n))$ levels of sampling. For each component in a given round, the total space required is $\widetilde{O}(r  / \eps^2)$ bits by \cref{thm:kpssketch}. Hence, the total space required here is $\widetilde{O}(nr / \eps^2)$ bits as well. 
	
	Finally, we must also be careful with the space requirements of storing the connected component information at different levels of sampling. Here we are simply storing the connectivity information of the hypergraph over $\log(m) \leq r \log(n)$ levels of sampling. Storing a partition of $n$ vertices requires $O(n \log(n))$ bits of space. Hence, across the $\leq r \log(n)$ levels of sampling, the total contribution is $O(nr \log^2(n))$ bits of space.
	
	Therefore, in total the space required by the sketch is $\widetilde{O}(nr / \eps^2)$ bits, as we desire. 
\end{proof}

Finally, we provide a proof of correctness; namely that the above algorithm does successfully create a $(1 \pm \eps)$ cut-sparsifier with high probability. We use this auxiliary claim:

\begin{claim}\label{clm:recoversLowStrength}
	Let $H_i$ denote the hypergraph at level $i$ of the recovery of \cref{alg:recoverSparsifier}. Then, with probability $1 - 1 / \mathrm{poly}(n)$, opening the sketches $\mathcal{S}_{C, i}: C \in \mathrm{CC}^{(i + 20 \log(n))}$ recovers exactly the hyperedges in $H_i$ of strength $\leq \kappa =  \frac{C \log^5(n / \eps)}{ \eps^2}$.
	\end{claim}

\begin{proof}
First, observe that the procedure for calculating connected components in \cref{alg:buildingSketch} is exactly the same as in \cref{alg:findStrongComp}. Indeed, each hyperedge $e$ survives sampling to level $i$ with probability $1 / 2^i$, and once at this level, any connected components that touch $e$ are merged.

Next, we will proceed by induction. We suppose the above claim is true up to level $i-1$. Now, observe that if this is the case, then our recovery procedure has exactly simulated \cref{alg:simpleSparsify} up to level $i-1$, and thus $H_i$ is distributed exactly as in \cref{alg:simpleSparsify}.  By \cref{cor:boundedHyperedges}, we then see that with high probability, $H_i /  \mathrm{CC}^{(i + 20 \log(n))}$ has $\leq n^{101}$ hyperedges, and any hyperedge of strength $\leq n^{15}$ in $H_i$ is still in $H_i /  \mathrm{CC}^{(i + 20 \log(n))}$, where $\mathrm{CC}^{(i + 20 \log(n))}$ are the connected components resulting from the connected components calculated in \cref{alg:findStrongComp} (which is distributed the same as the components found by \cref{alg:buildingSketch}). By construction, we are also in possession of the linear sketch of \cref{thm:kpssketch} on the set of components $\mathrm{CC}^{(i + 20 \log(n))}$, with parameter $\kappa =  \frac{C \log^5(n / \eps)}{ \eps^2} $. It follows then that when we open the sketch (to recover hyperedges), with all but $1  / \mathrm{poly}(n)$ probability, this recoveres exactly all edges of strength $\leq \kappa$.

The high probability in the above bound follows from the fact that there are $O(r \log(n))$ levels of recovery, and each one succeeds with $1 -1 / \mathrm{poly}(n)$ probability. 
\end{proof}

Now, we are able to conclude the proof of the accuracy:

\begin{claim}\label{clm:accuracyInsertion}
	For a hypergraph $H$ of arity $\leq r$ given through an insertion-only stream, \cref{alg:recoverSparsifier} recovers a $(1 \pm \eps)$ cut-sparsifier of $H$ with high probability. 
\end{claim}

\begin{proof}
Indeed, by the previous claim, we know that with probability $1 - 1 / \mathrm{poly}(n)$ (over the randomness of the linear sketches), the sketches in level $i$ of \cref{alg:buildingSketch} recover exactly the edges of strength $\leq \kappa = \kappa =  \frac{C \log^5(n / \eps)}{ \eps^2}$ in $H_i$. It follows then that this recover exactly performs the recovery of \cref{alg:simpleSparsify}, with a paramter $\eps' = \eps / \log^2(n / \eps)$. Thus, by \cref{thm:kpsaccuracy}, this yields a $(1 \pm \eps)$ cut-sparsifier with high probability. 
\end{proof}

We then conclude this section:

\begin{proof}[Proof of \cref{thm:insertionOnly}]
The space required for the sketch follows from \cref{alg:buildingSketch}, and \cref{clm:sketchSpaceInsertion}. The accuracy of the recovered sparsifier follows from \cref{alg:recoverSparsifier} and \cref{clm:accuracyInsertion}.
\end{proof}

In the next section, we discuss how a natural generalization of these ideas can be used to handle \emph{bounded deletion streams} without venturing entirely into the linear sketching regime.

\section{Bounded-deletion Streaming Algorithms}\label{sec:booundedDeletion}

In this section, we will show the following:

\begin{theorem}\label{thm:boundedDeletion}
	For $k \geq 1$, there is a $k$-bounded deletion algorithm requiring $\widetilde{O}(n r \log(k) / \eps^2)$ bits of space which creates a $(1 \pm \eps)$ cut-sparsifier for a hypergraph on $n$ vertices and hyperedges of arity $\leq r$, with probability $1 - 1 / \mathrm{poly}(n)$.
\end{theorem}

\begin{remark}
	One way to interpret the above result is as a smooth interpolation between the insertion-only and linear sketching modes. Linear sketching corresponds with the setting where \emph{every} hyperedge can be deleted, leading to $k = m$. On the other hand, insertion-only streams correspond with when $k = 0$.
\end{remark}

We present the algorithm for achieving this in the next subsection.

\subsection{Bounded-deletion Strong Component Implementation}

Let us recall \cref{alg:buildingSketch}. The key improvement in this algorithm was in observing that once a component has been designated as \emph{strongly connected}, this component will never cease to be strongly connected. This is because in that setting, the stream only allows insertions. Clearly, in the current setting, such an argument is not valid, as even with bounded deletions, it is certainly possible that a component will become less strong. However, the algorithm can be extended in a simple way which solves this short-coming: indeed, instead of merging components once their strength reaches some $\mathrm{poly}(n)$ level, we instead merge them when the strength reaches $\mathrm{poly}(k \cdot n)$. For instance, if a component has strength $(kn)^2$, then even after some $k$ deletions, the strength of this component will remain large. We formalize this below:

\begin{claim}\label{clm:deleteStrength}
	Let $H$ be a hypergraph, and let $C \subseteq V$ denote some set of vertices in $H$. Let $T \subseteq E$ denote some set of $\leq k$ hyperedges to be deleted. Then,
	\[
	\lambda_{H-T}(C) \geq \lambda_H(C) - k.
	\]
	That is, the strength of $C$ in $H-T$ decreases by at most $k$ compared to the strength of $C$ in $H$.
\end{claim}

\begin{proof}
Recall that the strength of a component $C$ is equal to the minimum normalized $k$-cut in the induced sub-graph $H[C]$. That is,
\[
\lambda_H(C) = \Phi(H[C]) = \min_{k} \min_{V_1, \dots V_k \in \mathcal{P}(C)} \frac{|E_{H[C]}[V_1, \dots V_k]|}{k-1}.
\]
Now, after deleting $k$ hyperedges, observe that the numerator decreases by at most $k$. Since the denominator is always $\geq 1$, this means that deleting $k$ hyperedges can only decrease the strength by $\leq k$. This yields the proof. 
\end{proof}

Now, we are ready to present the construction of our bounded-deletion sketch. We first re-present the estimation of strong components first, as a separate algorithm:

\begin{algorithm}[H]
	\caption{FindStrongComponents$(H)$}\label{alg:findStrongCompBounded}
	$\widetilde{H}^{(0)} = H$. \\
	\For{$i = 1, \dots r \log(n)$}
	{
		Let $\mathrm{CC}^{(i)}$ be the connected component structure of $\widetilde{H}^{(i)}$. \\
		Let $\widetilde{H}^{(i+1)} = \widetilde{H}^{(i)}$ with edges sampled at rate $1/2$. \\
	}
	\Return{$\{\mathrm{CC}^{(i)}: i \in [r \log(n)] \}$}
\end{algorithm}

Naively, if we wanted to build a sparsification scheme resilient to $k$ deletions, we could merge components in $H_i$ in correspondence with the components $\mathrm{CC}^{(i + 20 \log(n) + 2 \log(k))}$. This way, because the components would have strength $\mathrm{poly}(nk)$, by \cref{clm:deleteStrength}, they would remain strong even after $k$ deletions. While this does yield a non-trivial result, it does not directly solve our problem. Indeed, the analog of \cref{clm:numberLevelsInsertion} would imply that a component only has a non-empty neighborhood for $O(\log(n) + \log(k))$ levels of sampling. However, we must \emph{also} pay this factor of $\log(k)$ in the support size of the hypergraph as per \cref{thm:kpssketch} (as $m$ is now as large as $\mathrm{poly}(k)$). This would lead to a dependence of $\log(k)^2$, which is too large for us. 

Instead, we augment \cref{alg:findStrongCompBounded} with a linear-sketch based version, which provides refined information about the connected components. We state this algorithm below:

\begin{algorithm}[H]
	\caption{BoundedDeletionStrongComponents($H, \eps)$)}\label{alg:buildingComponentSketchBounded}
	To start, initialize $\mathrm{CC}^{(i)} = [n]$ for each $i$. \\
	Initialize a connected component linear sketch for each vertex $v \in V$, for each $i \in [r \log(n)]$ as per \cref{thm:gmtsketch} for support size $m \leq \mathrm{poly}(n)$.  Denote the sketch for vertex $v$, level $i$, by $\mathcal{S}^{(\mathrm{conn})}_{v,i}$. \\
	\For{$(e, \delta_e = \pm1))$ arriving in the stream}{
		\For{$i \in \{0, 1, \dots  r \log(n)\}$}{
			Update $\mathrm{CC}^{(i)}$ to merge any components connected by $e$. \\
			If any components $C_1, \dots C_{\ell}$ are merged together in $\mathrm{CC}^{(i)}$  
			to create a larger strong component $C' = C_1 \cup C_2 \cup \dots C_{\ell}$, 
			set $\mathcal{S}^{(\mathrm{conn})}_{C', i - 20 \log(nk)} = \mathcal{S}^{(\mathrm{conn})}_{C_1, i - 20 \log(nk)} + \mathcal{S}^{(\mathrm{conn})}_{C_2, i - 20 \log(nk)}  + \dots + \mathcal{S}^{(\mathrm{conn})}_{C_{\ell}, i - 20 \log(nk)} $, and delete $\mathcal{S}^{(\mathrm{conn})}_{C_1, i - 20 \log(nk)}, \dots \mathcal{S}^{(\mathrm{conn})}_{C_{\ell}, i - 20 \log(nk)}$. \\
			With probability $1/2$; end \textbf{for}, otherwise, continue.
		}
		\For{$i \in \{0, 1, \dots r \log(n) \}$}{
			Add $\delta_e \cdot e$ to every linear sketch $\mathcal{S}^{(\mathrm{conn})}_{C, i}: C \in \mathrm{CC}^{(i + 20 \log(nk))}$. \\
			With probability $1/2$; end \textbf{for}, otherwise, continue.
			}
	}
\end{algorithm}

Now, we show how to recover strong components from this sketch:

\begin{algorithm}[H]
	\caption{RecoverComponents($H, \eps)$)}\label{alg:recoverStrongCompBounded}
	\For{$i \in \{r \log(n), r\log(n) - 1, \dots 1, 0\}$}{
		Open the linear sketches $\mathcal{S}^{(\mathrm{conn})}_{C, i}$ in accordance with \cref{thm:gmtsketch}. \\
		Denote the resulting connected components by $\widetilde{\mathrm{CC}}^{(i)}$. \\
		Merge the $\ell_0$-samplers at level $i - 1$ in accordance with the components $\widetilde{\mathrm{CC}}^{(i)}$.\\ 
	}
	\Return{$\{\widetilde{\mathrm{CC}}^{(i)}: i \in [r \log(n)]\}$.}
\end{algorithm}

There are several claims we would like to show about this algorithm. First, we would like to bound the space required by the implementation:

\begin{claim}\label{clm:numberLevelsBounded}
	Fix a timestep $t$. Let $S$ be a component which appears in $\mathrm{CC}^{(i)}$ for some $i \in [r \log(n)]$. Then, there are only $O(\log(nk))$ levels of sampling in \cref{alg:buildingComponentSketchBounded} where the neighborhood of $S$ is non-empty with probability $1 - 1 / n^{99}$.
\end{claim}

\begin{proof}
	Let $j$ denote the first round in which $S$ appears in $\mathrm{CC}^{(j)}$. Then, by \cref{cor:boundedHyperedges}, the number of hyperedges total in $H_{j - 20 \log(nk)} / \mathrm{CC}^{(j)}$ is at most $(nk)^{101}$. In particular, this means that the number of hyperedges which can be incident on $S$ in $H_{j - 20 \log(nk)} / \mathrm{CC}^{(j)}$ is also bounded by $(nk)^{101}$. Thus, with high probability, after say, $200 \log(nk)$ levels of sampling, all hyperedges incident on $S$ will have been removed, and hence its neighborhood is empty.
\end{proof}

Now, if the neighborhood of $S$ is empty, then to store $\ell_0$ samplers for this neighborhood is entirely trivial. Using this, we can bound the total space required by \cref{alg:buildingSketch}.

\begin{claim}\label{clm:sketchComponentsSpaceBounded}
	At any timestep $t$, the total space required by \cref{alg:buildingComponentSketchBounded} is $\widetilde{O}(nr \log(k) / \eps^2)$ bits with probability $1 - 1 / \mathrm{poly}(n)$.
\end{claim}

\begin{proof}
	First, for at most $\log(m) \leq r \log(n)$ levels, we store $1$ bit for $\leq n$ vertices' empty $\ell_0$ samplers. Thus, the total contribution from these empty samplers is $\widetilde{O}(nr)$ bits. 
	
	Now, for the non-empty $\ell_0$ samplers, we must only store these for $O(n)$ strong connected components (by \cref{clm:numberLevelsInsertion}), for $O(\log(nk))$ levels of sampling. For each component in a given round, the total space required is $\widetilde{O}(r)$ bits by \cref{thm:gmtsketch}. Hence, the total space required here is $\widetilde{O}(nr \log(k))$ bits. 
	
	Finally, we must also be careful with the space requirements of the pre-processing phase. Here we are simply storing the connectivity information of the hypergraph over $\log(m) \leq r \log(n)$ levels of sampling. Storing a partition of $n$ vertices requires $O(n \log(n))$ bits of space. Hence, across the $\leq r \log(n)$ levels of sampling, the total contribution is $O(nr \log^2(n))$ bits of space.
	
	Therefore, in total the space required by the sketch is $\widetilde{O}(nr \log(k))$ bits, as we desire. 
\end{proof}

First, we observe that we have the following:

\begin{claim}
		Let $H$ be a fixed hypergraph, and let $i \in [r \log(n)]$. Let $H_i$ be hypergraph resulting from \cref{alg:simpleSparsify} at the $i$th level of sampling. Then, the hypergraph $H_i /  \mathrm{CC}^{(i + 20 \log(nk))}$ has $\leq (nk)^{101}$ hyperedges, and any hyperedge of strength $\leq (nk)^{15}$ in $H_i$ is still in $H_i /  \mathrm{CC}^{(i + 20 \log(nk))}$.
\end{claim}

\begin{proof}
This follows from \cref{cor:boundedHyperedges}.
\end{proof}

Next, using this we immediately obtain the following:

\begin{claim}\label{clm:respectLowStrength}
	Let $H$ be a hypergraph resulting from a $k$-bounded deletion stream, and let $i \in [r \log(n)]$. Let $H_i$ be hypergraph resulting from \cref{alg:simpleSparsify} at the $i$th level of sampling. Then, the hypergraph $H_i /  \mathrm{CC}^{(i + 20 \log(nk))}$ has $\leq (nk)^{101}$ hyperedges, and any hyperedge of strength $\leq (nk)^{15} - k$ in $H_i$ is still in $H_i /  \mathrm{CC}^{(i + 20 \log(nk))}$.
\end{claim}

\begin{proof}
First, the reason the above claim is not trivial is that the connected components $\mathrm{CC}^{(i)}$ that are estimated do not account for deletions in the stream. Instead, these are computed with respect to the hypergraph $\hat{H}$, which is all hyperedges that have been inserted. With respect to $\hat{H}$ (and running \cref{alg:simpleSparsify} on $\hat{H}$), it is the case that (with high probability) the hypergraph $\hat{H}_i /  \mathrm{CC}^{(i + 20 \log(nk))}$ has $\leq (nk)^{101}$ hyperedges, and any hyperedge of strength $\leq (nk)^{15}$ in $\hat{H}_i$ is still in $\hat{H}_i /  \mathrm{CC}^{(i + 20 \log(nk))}$ by \cref{cor:boundedHyperedges}. Observe that initially, every connected component in $\mathrm{CC}^{(i + 20 \log(nk))}$ was of strength $\geq (nk)^{15}$ in $\hat{H}_i$, as per \cref{clm:subsamplingGivesStrong}. Thus, after deleting $k$ hyperedges to get $H$ (and the sequence $H_i$), by \cref{clm:deleteStrength}, these same components $\mathrm{CC}^{(i + 20 \log(nk))}$ have strength greater than $(nk)^{15} - k$ in $H_i /\mathrm{CC}^{(i + 20 \log(nk))}$. Thus, any hyperedge of strength $\leq (nk)^{15} - k$ is still in $H_i /  \mathrm{CC}^{(i + 20 \log(nk))}$.

For the upper bound on the number of hyperedges, this follows trivially by \cref{cor:boundedHyperedges}, as deleting hyperedges can only decrease the number of remaining hyperedges. This yields the claim.
\end{proof}

Note that in \cref{alg:recoverStrongCompBounded}, there is component information coming from two sources: on the one hand, we are using the pre-processing phase to roughly identify strong components on the order of $\mathrm{poly}(nk)$. During recovery time, we recover the connectivity information \emph{from the bottom-up}, starting with the most aggressive sampling rate. This is because if the hypergraph is connected at a much smaller sampling rate, then it will also be connected at a slightly higher sampling rate. The reason we do this recovery process in reverse is so that we can ensure that (after merging components) the number of hyperedges in the hypergraph is bounded. 

Before proceeding, we first introduce a piece of notation: in \cref{alg:buildingComponentSketchBounded}, let $H'_i$ refer to the sequence of hypergraphs that are constructed by sub-sampling in Lines 9-13. We show the following claim:

\begin{claim}\label{clm:componentRecovery}
	With probability $1 - 1  / \mathrm{poly}(n)$, the components $\{\widetilde{\mathrm{CC}}^{(i)}: i \in [r \log(n)]\}$, as returned by \cref{alg:recoverStrongCompBounded}, are exactly the connected components of $H'_i$.
\end{claim}

\begin{proof}
First, by \cref{clm:respectLowStrength}, it is the case that the components $\mathrm{CC}^{(i + 20 \log(nk))}$ do not merge components that were not already connected. That is to say, the connectivity structure that results from $H'_i / \mathrm{CC}^{(i + 20 \log(nk))}$ is the same as the structure that exists in $H'_i$. Next, we observe that because the sketch in \cref{alg:recoverStrongCompBounded} is opened in reverse (from the most aggressive sampling rate), is thus also the case that $H'_i / \mathrm{CC}^{(i + 20 \log(nk))} / \{\widetilde{\mathrm{CC}}^{(i+1)}\}$ has the same connectivity structure as $H'_i / \mathrm{CC}^{(i + 20 \log(nk))}$, as any component that is connected at the more aggressive sampling rate is also connected at the less aggressive sampling rate.

Now, it remains only to show that the $\ell_0$ sampler sketch from \cref{thm:gmtsketch} succeeds in recovering the connectivity structure of $H'_i / \mathrm{CC}^{(i + 20 \log(nk))} / \{\widetilde{\mathrm{CC}}^{(i+1)}$. For this, the only condition we must show is that the support size of this resulting contracted hypergraph is bounded. However, this follows very easily: 

\begin{claim}
	Let $G$ be a hypergraph, and let $G'$ be the result of sampling $G$ at rate $1/2$. Let $\mathrm{CC}(G')$ denote the connected components of $G'$. Then, $G / \mathrm{CC}(G')$ has at most $n^5$ hyperedges with probability $1 - 1 / 2^{\Omega(n)}$.
\end{claim}

\begin{proof}
Fix any partition $Q$ of the vertex set $V$. Observe that if $|E_G(Q)| \geq n^5$, then $|E_{G'}(Q)| \geq 1$ with probability $1 - 1 / 2^{n^5}$. Taking a union bound over all $\leq n^n$ possible partitions, we see that with probability $1 - 1 / 2^n$, there is no partition $Q$ such that $|E_{G'}(Q)| = 0$, but $|E_G(Q)| \geq n^5$. This yields the claim. 
\end{proof}

By invoking this claim with the hypergraph $H'_i / \mathrm{CC}^{(i + 20 \log(nk))} / \{\widetilde{\mathrm{CC}}^{(i+1)} \}$, this proves that with all but exponentially small probability, the support of the hypergraph is bounded by $\mathrm{poly}(n)$. Thus, the $\ell_0$-samplers can be defined with parameter $m \leq \mathrm{poly}(n)$, while still producing the correct connectivity structure with probability $1 - 1 / \mathrm{poly}(n)$. 
\end{proof}

Now, we can prove the following:

\begin{claim}\label{clm:boundedHyperedgesBoundedDeletion}
	Let $H$ be a hypergraph, and let $i \in [r \log(n)]$. Let $H_i$ be hypergraph resulting from \cref{alg:simpleSparsify} at the $i$th level of sampling. Then, with high probability over the output $\{\widetilde{\mathrm{CC}}^{(i)}: i \in [r \log(n)]\}$ from \cref{alg:recoverStrongCompBounded}, the hypergraph $H_i /  \widetilde{\mathrm{CC}}^{(i + 20 \log(n))}$ has $\leq n^{101}$ hyperedges, and any hyperedge of strength $\leq n^{15}$ in $H_i$ is still in $H_i /  \widetilde{\mathrm{CC}}^{(i + 20 \log(n))}$.
\end{claim}

\begin{proof}
	By \cref{clm:componentRecovery}, the components recovered in \cref{alg:recoverStrongCompBounded} are exactly the components in $H'_i$. $H'_i$ is simply a sequence of sub-sampled hypergraphs, and thus the distribution over its components is exactly the same as yielded by \cref{alg:findStrongComp}. But for these components, we already know that by \cref{cor:boundedHyperedges}, the stated claim is true.
\end{proof}

Note that the key benefit here is that the number of hyperedges no longer has a dependence on $k$.

\subsection{Bounded-deletion Sparsification Implementation}

Now that we have established some properties about the strong-component implementation, we are ready to proceed to the actual implementation of our sparsification scheme. We present the algorithm below:

\begin{algorithm}[H]
	\caption{StreamingAlgorithm($H, \eps)$)}\label{alg:buildingSketchBounded}
	To start, initialize $\mathrm{CC}^{(i)} = [n]$ for each $i$. \\
	Initialize the sketch of \cref{thm:kpssketch} for each vertex $v \in V$, for each $i \in [r \log(n)]$ using $\kappa  = \frac{C \log^5(n / \eps)}{ \eps^2}$, $m = \mathrm{poly}(n)$. Denote the sketch for vertex $v$, level $i$, by $\mathcal{S}_{v,i}$. \\
	\For{$(e, \delta_e = \pm 1))$ arriving in the stream}{
		\For{$i \in \{0, 1, \dots  r \log(n)\}$}{
			Update $\mathrm{CC}^{(i)}$ to merge any components connected by $e$. \\
			If any components $C_1, \dots C_{\ell}$ are merged together in $\mathrm{CC}^{(i)}$  to create a larger strong component $C' = C_1 \cup C_2 \cup \dots C_{\ell}$, set $\mathcal{S}_{C', i - 20 \log(n)} = \mathcal{S}_{C_1, i - 20 \log(n)} + \mathcal{S}_{C_2, i - 20 \log(n)}  + \dots + \mathcal{S}_{C_{\ell}, i - 20 \log(n)} $, and delete $\mathcal{S}_{C_1, i - 20 \log(n)}, \dots \mathcal{S}_{C_{\ell}, i - 20 \log(n)}$. \\
			With probability $1/2$; end \textbf{for}, otherwise, continue.
		}
		\For{$i \in \{0, 1, \dots r \log(n) \}$}{
			Add $\delta_e \cdot e$ to every linear sketch $\mathcal{S}^{(\mathrm{conn})}_{C, i}: C \in \mathrm{CC}^{(i + 20 \log(nk))}$. \\
			With probability $1/2$; end \textbf{for}, otherwise, continue.
		}
		\For{$i \in \{0, 1, \dots r \log(n)\}$}{
		Add $e\cdot \delta_e$ to every linear sketch $\mathcal{S}_{C, i}: C \in \mathrm{CC}^{(i + 20 \log(n))}$. \\
		With probability $1/2$; end \textbf{for}, otherwise, continue.}
	}
\end{algorithm}

We also present the recovery algorithm that we use to find the $(1 \pm \eps)$ cut-sparsifier:

\begin{algorithm}[H]
	\caption{RecoverComponents($H, \eps)$)}\label{alg:recoverSparsifierBounded}
	\For{$i \in \{r \log(n), r\log(n) - 1, \dots 1, 0\}$}{
		Open the linear sketches $\mathcal{S}^{(\mathrm{conn})}_{C, i}$ in accordance with \cref{thm:gmtsketch}. \\
		Denote the resulting connected components by $\widetilde{\mathrm{CC}}^{(i)}$. \\
		Merge the $\ell_0$-samplers at level $i - 1$ in accordance with the components $\widetilde{\mathrm{CC}}^{(i)}$.\\ 
	}
	\For{$i \in \{0, 1, \dots  r \log(n)\}$}{
		Open the linear sketches $\mathcal{S}_{C, i}$ in accordance with \cref{thm:kpssketch}. \\
		Denote the recovered hyperedges by $F_i$. \\
		Remove these recovered hyperedges from sketches at levels $j > i$. 
	}
	\Return{$\bigcup_i 2^i \cdot F_i$.}
\end{algorithm}

\subsection{Bounded-deletion Sparsification Analysis}

As before, we first bound the space required by these algorithms:

\begin{claim}\label{clm:sketchSpaceBounded}
	At any timestep $t$, the total space required by \cref{alg:buildingSketchBounded} is $\widetilde{O}(nr\log(k) / \eps^2)$ bits with probability $1 - 1 / \mathrm{poly}(n)$.
\end{claim}

\begin{proof}
	First,  by \cref{clm:sketchComponentsSpaceBounded}, the space required to store the connectivity sketches is $\widetilde{O}(n r \log(k))$ bits. Thus, we focus only on bounding the space required for the sketches of \cref{thm:kpssketch}. For this, recall that for at most $\log(m) \leq r \log(n)$ levels, we store $1$ bit for $\leq n$ vertices' empty $\ell_0$ samplers (initialized for a support of size $\mathrm{poly}(n)$. Thus, the total contribution from these empty samplers is $\widetilde{O}(nr)$ bits. 
	
	Now, for the non-empty $\ell_0$ samplers, we must only store these for $O(n)$ strong connected components, for $O(\log(nk))$ levels (by \cref{clm:numberLevelsBounded}). For each component in a given round, the total space required is $\widetilde{O}(r  / \eps^2)$ bits by \cref{thm:kpssketch} (using support $\mathrm{poly}(n)$, and $\kappa = \polylog(n / \eps)$. Hence, the total space required here is $\widetilde{O}(nr \log(k)/ \eps^2)$ bits, and thus the total space required is $\widetilde{O}(nr \log(k)/ \eps^2)$ bits.
\end{proof}

Finally, we provide a proof of correctness; namely that the above algorithm does successfully create a $(1 \pm \eps)$ cut-sparsifier with high probability. We use this auxiliary claim:

\begin{claim}\label{clm:recoversLowStrengthBounded}
	Let $H_i$ denote the hypergraph at level $i$ of the recovery of \cref{alg:recoverSparsifierBounded}. Then, with probability $1 - 1 / \mathrm{poly}(n)$, opening the sketches $\mathcal{S}_{C, i}: C \in \widetilde{\mathrm{CC}}^{(i + 20 \log(n))}$ recovers exactly the hyperedges in $H_i$ of strength $\leq \kappa =  \frac{C \log^5(n / \eps)}{ \eps^2}$.
\end{claim}

\begin{proof}
	First, observe that the procedure for calculating connected components in \cref{alg:buildingSketchBounded} is exactly the same as in \cref{alg:findStrongComp}. Indeed, each hyperedge $e$ survives sampling to level $i$ with probability $1 / 2^i$, and once at this level, any connected components that touch $e$ are merged.

	Next, we will proceed by induction. We suppose the above claim is true up to level $i-1$. Now, observe that if this is the case, then our recovery procedure has exactly simulated \cref{alg:simpleSparsify} up to level $i-1$, and thus $H_i$ is distributed exactly as in \cref{alg:simpleSparsify}.  By \cref{clm:boundedHyperedgesBoundedDeletion}, we then see that with high probability, $H_i /  \widetilde{\mathrm{CC}}^{(i + 20 \log(n))}$ has $\leq n^{101}$ hyperedges, and any hyperedge of strength $\leq n^{15}$ in $H_i$ is still in $H_i /  \widetilde{\mathrm{CC}}^{(i + 20 \log(n))}$, where $\widetilde{\mathrm{CC}}^{(i + 20 \log(n))}$ are the connected components resulting from the connected components calculated in \cref{alg:findStrongComp} (which is distributed the same as the components found by \cref{alg:buildingSketch}). By construction, we are also in possession of the linear sketch of \cref{thm:kpssketch} on the set of components $\widetilde{\mathrm{CC}}^{(i + 20 \log(n))}$, with parameter $\kappa =  \frac{C \log^5(n / \eps)}{ \eps^2} $. It follows then that when we open the sketch (to recover hyperedges), with all but $1  / \mathrm{poly}(n)$ probability, this recovers exactly all edges of strength $\leq \kappa$.
	
	The high probability in the above bound follows from the fact that there are $O(r \log(n))$ levels of recovery, and each one succeeds with $1 -1 / \mathrm{poly}(n)$ probability. 
\end{proof}

Now, we are able to conclude the proof of the accuracy:

\begin{claim}\label{clm:accuracyBounded}
	For a hypergraph $H$ of arity $\leq r$ given through a $k$-bounded deletion stream, \cref{alg:recoverSparsifier} recovers a $(1 \pm \eps)$ cut-sparsifier of $H$ with high probability. 
\end{claim}

\begin{proof}
	Indeed, by the previous claim, we know that with probability $1 - 1 / \mathrm{poly}(n)$ (over the randomness of the linear sketches), the sketches in level $i$ of \cref{alg:buildingSketchBounded} recover exactly the edges of strength $\leq \kappa = \kappa =  \frac{C \log^5(n / \eps)}{ \eps^2}$ in $H_i$. It follows then that this algorithm exactly performs the recovery of \cref{alg:simpleSparsify}, with a parameter $\eps' = \eps / \log^2(n / \eps)$. Thus, by \cref{thm:kpsaccuracy}, this yields a $(1 \pm \eps)$ cut-sparsifier with high probability. 
\end{proof}

We then conclude this section:

\begin{proof}[Proof of \cref{thm:boundedDeletion}]
	The space required for the sketch follows from \cref{alg:buildingSketchBounded}, and \cref{clm:sketchSpaceBounded}. The accuracy of the recovered sparsifier follows from \cref{alg:recoverSparsifierBounded} and \cref{clm:accuracyBounded}.
\end{proof}

In the next section, we complement this algorithm with a nearly-matching lower bound.

\section{Lower Bounds for Creating Sparsifiers}\label{sec:lowerbound}

First, we recall the Augmented Index problem (denoted $\textbf{IND}_N$):

\begin{enumerate}
    \item Alice is given a vector $y \in \zo^N$, and sends a message $M$ to Bob.
    \item Bob is given an index $i^* \in [N]$, $y_{i^* + 1}, \dots y_N$, and the message $M$, and must output the bit $y_{i^*}$ with probability $\geq 2/3$. 
\end{enumerate}

The following is known regarding $\textbf{IND}_N$:

\begin{fact}\cite{CK11}
    Any protocol which solves $\textbf{IND}_N$ must use $|M| = \Omega(n)$.
\end{fact}

In particular, the proof of the above fact is entirely information theoretic (in fact, \cite{CK11} shows that a specific quantity called the \emph{information cost} is lower bounded). Thus, we immediately have the following: 

\begin{definition}
    Let $\ell, N$ be integers. We define the $\ell$-Augmented index problem $\ell-\textbf{IND}_N$ to be $\ell$ simultaneous instances of the augmented index problem:
    \begin{enumerate}
        \item Alice is given $\ell$ vectors $y^{(1)}, \dots y^{(\ell)} \in \zo^N$, and sends a message $M$ to Bob.
    \item Bob is given $\ell$ indices $i_1^*, \dots i_{\ell}^* \in [N]$. Bob is also given the message $M$, and for each $j \in [\ell]$, Bob is given $y^{(j)}_{i_j^* + 1}, \dots y^{(j)}_N$. Bob must output $y^{(j)}_{i_{j}^*}$ for each $j \in [\ell]$ with probability $\geq 2/3$. 
    \end{enumerate}
\end{definition}

Naturally, we then have the following:

\begin{corollary}\label{cor:LBManyAugIndex}
    Any protocol which solves $\ell-\textbf{IND}_N$ must use $|M| = \Omega(n\ell)$.\footnote{Technically, we must bound the information cost, but this follows exactly from \cite{CK11}.}
\end{corollary}

Now, we introduce the notion of a support-sampler for a bounded deletion stream:

\begin{definition}
    Let $f \in \zo^N$ be constructed by a $k$-bounded deletion stream, where each update of the stream specifies $f_i + 1$ or $f_i - 1$. A one-pass support sampler is any algorithm that gets one pass through the stream and must output any $i \in [N]$ such that $f_i \neq 0$.
\end{definition}

Next, we recall the following theorem from \cite{JW18}:

\begin{theorem}[Theorem 20 of \cite{JW18}]\label{thm:JW18LB}
    Any one-pass support sampler which outputs an arbitrary $i \in [N]$ such that $f_i \neq 0$ with probability $\geq 2/3$ can be used to solve $\textbf{IND}_{p}$, for some $p = \Omega(\log(N/k)\log(k))$.
\end{theorem}

With this, we can easily derive a corollary for the case of solving $\ell$-$\textbf{IND}$:

\begin{corollary}
    Any streaming algorithm which (simultaneously) performs support-sampling on $\ell$ streams, each on a universe of $N$ items, with $\leq k$-deletions, and overall success probability $\geq 2/3$, must use $\Omega(\ell \log(N/k)\log(k))$ bits of memory.
\end{corollary}

\begin{proof}
    To start, let us suppose we are given some instance $I$ of $\ell-\textbf{IND}_{p}$, for some $p = \Omega(\log(N/k) \log(k)$ as specified by \cref{thm:JW18LB}. Let us denote the sub-instances by $I_1, \dots I_{\ell}$. For each such instance $I_j$, by \cref{thm:JW18LB}, it follows that we can construct a stream $f^{(j)}$ over the universe $[N]$ such that one-pass support sampling with $\leq k$ deletions on $f^{(j)}$ solves the augmented index problem corresponding to $I_j$. Thus, because our algorithm at hand can (simultaneously) perform support-sampling on these $\ell$ streams with $\leq k$-deletions, and overall success probability $\geq 2/3$, this can be directly used to solve each instance $I_j$, and thus the overall instance $\ell-\textbf{IND}_{p}$. In particular, by \cref{cor:LBManyAugIndex}, this means that the algorithm at hand must use $\Omega(\ell \cdot p) = \Omega(\ell \log(N/k)\log(k))$ bits of memory.
\end{proof}

Finally, we are ready to conclude:

\begin{theorem}
    Any streaming algorithm for $k$-bounded deletion streams, which for hypergraphs on $n$ vertices, of arity $r \leq n/2+1$, produces a $(1 \pm \eps)$ cut-sparsifier for $\eps < 1$, must use $\Omega(n r \log(k/n))$ bits of space.
\end{theorem}

\begin{proof}
    We prove this by reducing the simultaneous support-sampling problem to the hypergraph sparsification problem. Indeed, suppose we are simultaneously given $\ell = n/2$ streams, each on a universe of size $N = 2^{r-1}$, with $\leq k/n$ deletions. Let us denote these streams by $f_1, \dots f_{\ell}$.

    We now create a hypergraph $H$ as follows: partition the vertex set $V$ into a left side $L$ and a right side $R$, each of size $n/2$. Now, construct a correspondence between the vertices in $L$ and the streams $f_1, \dots f_{\ell}$, such that vertex $i$ in $L$ corresponds with $f_i$. Next, for the universe $[2^{r-1}]$, we associate $2^{r-1}$ distinct subsets of $R$. Note that because $r-1 \leq n/2$, the number of subsets of $R$ is at least as large as $2^{r-1}$. We will abuse notation and for an element $x \in [2^{r-1}]$, we will also associate it with its corresponding subset $\subseteq R$.
    
    In stream $f_i$, whenever an element $x \in [2^{r-1}]$ is inserted or deleted, this then corresponds to inserting or deleting the hyperedge $\{ i \} \cup x$ into the hypergraph $H$. Note that because each of the $\ell$ streams has at most $k/n$ deletions, in total, we have that at most $k$ hyperedges are deleted. 

    Observe now that we have the following properties:
    \begin{enumerate}
        \item $H$ has $n$ vertices.
        \item The stream creating $H$ has at most $k$ deletions.
        \item The arity of $H$ is at most $r-1 + 1 = r$.
    \end{enumerate}
    In particular, this means that our streaming algorithm succeeds in creating a $(1 \pm \eps)$ cut-sparsifier of $H$ with probability $\geq 2/3$. We denote this sparsifier by $\hat{H}$. In particular, because $\eps <1$, this means that $\hat{H}$ must have the same connected component structure as $H$. 

    Now, we claim that we can use $\hat{H}$ to solve the support-sampling problem for each $f_i$. Indeed, consider any $f_i$. By definition, if $f_i$ had at least one element in its support (say element $x$), then $H$ has the hyperedge $\{ i \} \cup x$. Thus, in the sparsifier $\hat{H}$, there must also be some hyperedge recovered which is incident to the vertex $i$. But, every hyperedge incident to $i$ is of the form $\{i \} \cup x'$, for $x' \in [2^{r-1}]$. In particular, because $\hat{H}$ is a sparsifier of $H$, it must also be the case that every hyperedge in $\hat{H}$ is also in $H$. So, let us denote the recovered hyperedge by $\{i \} \cup x'$, for $x' \in [2^{r-1}]$. Because $\{i \} \cup x'$ is in $H$, this means that $x'$ is remaining in the stream $f_i$, and thus $x'$ is a valid element in the support of $f_i$.

    Hence, $\hat{H}$ can be used to recover a solution to each of the support-sampling problems for $f_1, \dots f_{n/2}$. In particular, by plugging in $\ell = n/2, N = 2^{r-1}$, and the number of deletions is $k/n$, we get a lower bound of $\Omega(n r \log(k/n))$ for the complexity of solving the support-sampling problem on these streams $f_1, \dots f_{n/2}$, and thus so too for the complexity of sparsifying hypergraphs in $k$-bounded deletion streams. This yields the theorem.
\end{proof}

 \bibliographystyle{alpha}
\bibliography{ref}
\appendix

\end{document}